\documentclass[12pt]{article}

\usepackage{longtable}
\usepackage{scicite}

\usepackage{times}
\usepackage{graphicx}
\usepackage{float}
\usepackage{subfigure}
\usepackage{setspace}

\usepackage{color}

\newenvironment{sciabstract}{%
\begin{quote} \bf}
{\end{quote}}

\newcounter{lastnote}

\newcommand{\wq}[1]{\textcolor{black}{$^{\textrm{}}${#1}}}

\title{Debunking in a World of Tribes}

\author
{Fabiana Zollo$^{1}$, Alessandro Bessi$^{1,2}$, Michela Del Vicario$^{1}$, Antonio Scala$^{1,3,4}$,\\  Guido Caldarelli$^{1,3,4}$, Louis Shekhtman$^{5}$, Shlomo Havlin$^{5}$, Walter Quattrociocchi$^{1\ast}$ \\
\\
\normalsize{$^{1}$IMT Alti Studi Lucca, It}\\
\normalsize{$^{2}$IUSS Pavia, It}\\
\normalsize{$^{3}$ISC-CNR Rome, It}\\
\normalsize{$^{4}$LIMS, London, Uk}\\
\normalsize{$^{5}$Bar-Ilan University, Il}\\
\\
\normalsize{$^\ast$walterquattrociocchi@gmail.com}
}

\date{}

\begin{document} 


\baselineskip24pt


\maketitle 

\begin{singlespacing}


\begin{sciabstract}
Recently a simple military exercise on the Internet was perceived as the beginning of a new civil war in the US.
Social media aggregate people around common interests eliciting a collective framing of narratives and worldviews.
However, the wide availability of user-provided content and the direct path between producers and consumers of information often foster confusion about causations, encouraging mistrust, rumors, and even conspiracy thinking. In order to contrast such a trend attempts to \textit{debunk} are often undertaken. 
Here, we examine the effectiveness of debunking through a quantitative analysis of 54 million users over a time span of five years (Jan 2010, Dec 2014). 
In particular,  we compare how users interact with proven (scientific) and unsubstantiated (conspiracy-like) information on Facebook in the US.
Our findings confirm the existence of echo chambers where users interact primarily with either conspiracy-like or scientific pages. Both groups interact similarly with the information within their echo chamber. 
We examine 47,780 debunking posts and find that attempts at debunking are largely ineffective. For one, only a small fraction of usual consumers of unsubstantiated information interact with the posts. Furthermore, we show that those few are often the most committed conspiracy users and rather than internalizing debunking information, they often react to it negatively.  
Indeed, after interacting with debunking posts, users retain, or even increase, their engagement within the conspiracy echo chamber.
\end{sciabstract}

\section*{Introduction}

Misinformation on online social media is pervasive and represents one of the main threats to our society according to the World Economic Forum \cite{howell2013}. 
The diffusion of false rumors affects public perception of reality as well as the political debate. 
Links between vaccines and autism, the belief that 9/11 was an inside job, or the more recent case of Jade Helm 15 (a simple military exercise that was perceived as the imminent threat of the civil war in the US), are just few examples of the consistent body of the collective narratives grounded on unsubstantiated information.

Socio-technical systems and microblogging platforms such as Facebook and Twitter have created a direct path of content from producers to consumers, changing the way users become informed, debate ideas, and shape their worldviews \cite{brown2007word,Richard2004,QuattrociocchiCL11,Quattrociocchi2014,Kumar2010}. 
This scenario might foster confusion about causations regarding global and social issues and thus encouraging paranoia based on false rumors, and mistrust \cite{sunstein2009conspiracy}.
Recently, researches have shown \cite{bessi2014science,mocanu2014,bessi2014economy, bessi2015trend, zollo2015emotional, del2015echo} that continued exposure to unsubstantiated rumors may be a good proxy to detect gullibility -- i.e., jumping the credulity barrier accepting highly implausible theories -- on online social media.
Narratives, especially those grounded in conspiracy theories, play an important cognitive and social function in that they simplify causation. Indeed, they are formulated in a way that is able to reduce the complexity of reality and to tolerate a certain level of uncertainty \cite{byford2011conspiracy,finerumor,hogg2011extremism}.
However, conspiracy thinking creates or reflects a climate of disengagement from mainstream society and recommended practices \cite{betsch2013debunking}. 

To combat conspiracy thinking, several algorithmic-driven solutions have been proposed \cite{qazvinian2011rumor,ciampaglia2015computational,resnick2014rumorlens,gupta2014tweetcred,almansour2014model,ratkiewicz2011detecting}. Further, Google \cite{dong2015knowledge} is currently studying the {\em trustworthiness score} to rank queries' results.  Along the same lines, Facebook proposed a community driven approach where users can tag false contents. 
However, such practices are controversial since they raise issues regarding the free circulation of contents. 
Moreover, recent studies \cite{bessi2014science,mocanu2014,Nyhan01042014,bessi2015viral} have shown that social and cognitive factors, like confirmation bias and social reinforcement, play a fundamental role in shaping content selection criteria and thus most consumers of false content are unlikely to tag it.  

In this work, we perform a thorough quantitative analysis of 54 million US Facebook users and study how they consume scientific and conspiracy-like content. We identify two main categories of pages: conspiracy news -- i.e. pages promoting contents {\em neglected} by main stream media -- and science news. We categorize pages according to their contents and their self-description. Using an approach based on  \cite{bessi2014science,mocanu2014,bessi2014economy}, we further explore Facebook pages that are active in debunking conspiracy theses and determine the effectiveness of these efforts (see Methods Section for further details).

Notice that we do not focus on the quality of the information but rather on the possibility for verification. Indeed, it is easy for scientific news to identify the authors of the study, the university under which the study took place and if the paper underwent a peer review process. On the other hand, conspiracy-like content is difficult to verify because it is inherently based upon suspect information and is derived allegations and a belief in secrets from the public. 
Notice that a fundamental aspect of the conspiracy narrative is the lack of trust in mainstream media, official institutions, and scientific researchers.  
The self-description of many conspiracy pages on Facebook claim that they inform people about topics neglected by mainstream media and science. Pages like {\em I don't trust the government}, {\em Awakening America}, or {\em Awakened Citizen}, promote wide-ranging content  from aliens, chem-trails, to the causal relation between vaccinations and autism or homosexuality. 
Conversely, science news pages -- e.g., {\em Science}, {\em Science Daily}, {\em Nature} -- are active in diffusing posts about the most recent scientific advances.  
To our knowledge, the final dataset is the complete set of all scientific and conspiracy-like information sources active in the US Facebook community and includes 83 and 330 pages, respectively. In addition, we identified 65 Facebook pages that focus on debunking conspiracy theories.

Our analysis reveals that two well-formed and highly segregated communities exist around conspiracy and scientific topics -- i.e., users are mainly active in only one category.  Focusing on users interactions with respect to their preferred content, we find similarities in the way the both forms of content is consumed. 

To determine whether online debunking campaigns against false rumors are effective, we measure the response of consumers of conspiracy stories to 47,780 debunking posts. 
We find that only a small fraction of users interact with debunking posts and that, when they do, their interaction often leads to increasing interest in conspiracy-like contents. 
These findings suggest that the problem behind misinformation is not only gullibility, but also conservatism. When users are confronted with new and untrusted opposing sources online, the interaction leads them to further commit to their own echo chamber.

\section*{Results and Discussion}

The aim of this work is to test the effectiveness of debunking campaigns in online social media. 
To do this, we start our analysis by statistically characterizing users attention' patterns with respect to unverified content and we use scientific news as a control. 
We then measure the effects of interaction by usual consumers of unsubstantiated claims with information aimed at correcting these false rumors -- i.e., debunking posts.

\subsection*{Echo chambers}

As a first step we characterize how distinct types of information -- belonging to the two different narratives -- are consumed on Facebook. In particular we focus on users' actions allowed by Facebook's interaction paradigm -- i.e., likes, shares, and comments. 
Each action has a particular meaning \cite{Ellison2007}. A {\em like} represents positive feedback to a post; a {\em share} expresses a desire to increase the visibility of given information; and a {\em comment} is the way in which online collective debates take form around the topic of the post. 
Therefore, comments may contain negative or positive feedbacks with respect to the post.

\begin{figure}[H]
	\centering
	\includegraphics[width=0.9\textwidth]{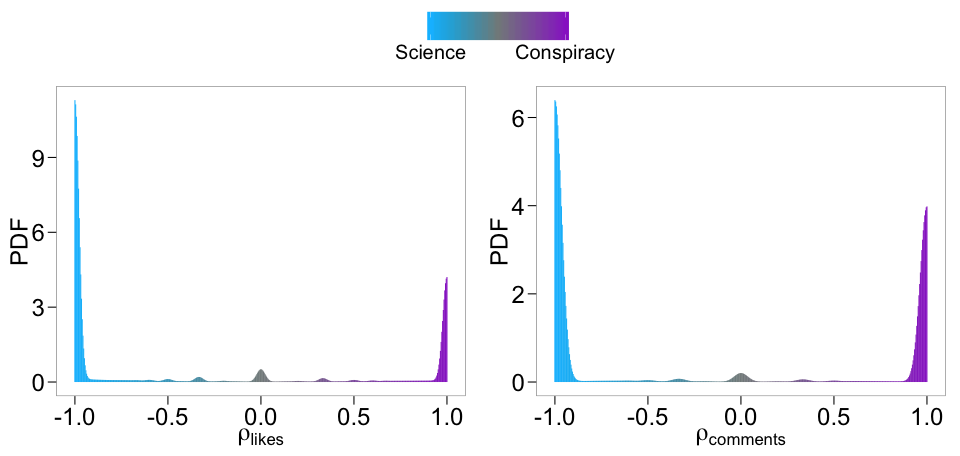} 
	\caption{\textbf{Users polarization.} Probability density functions (PDFs) of the polarization of all users computed on likes \textit{(left)} and comments \textit{(right)}.}
	\label{fig:pol_like}
\end{figure}

Assuming that user $u$ has performed $x$ and $y$ likes on scientific and conspiracy-like posts, respectively, we let $\rho(u)= (y-x)/(y+x)$. Thus, a user $u$ for whom $\rho(u)=-1$ is polarized towards science, whereas a user whose $\rho(u)=1$ is polarized towards conspiracy.
We define the user polarization $\rho \in [-1,1]$ as the ratio of difference in likes (or comments)  on conspiracy and science posts.  In Figure~\ref{fig:pol_like} we show that the probability density function (PDF) for the polarization of all the users is sharply bimodal with most having ($\rho(u) \sim -1$) or ($\rho(u) \sim 1$). Thus, most users may be divided into two groups, those \emph{polarized towards science} and those \emph{polarized towards conspiracy}. The same pattern holds if we look at polarization based on comments.

\begin{figure}[H]
	\centering
	\includegraphics[width=\textwidth]{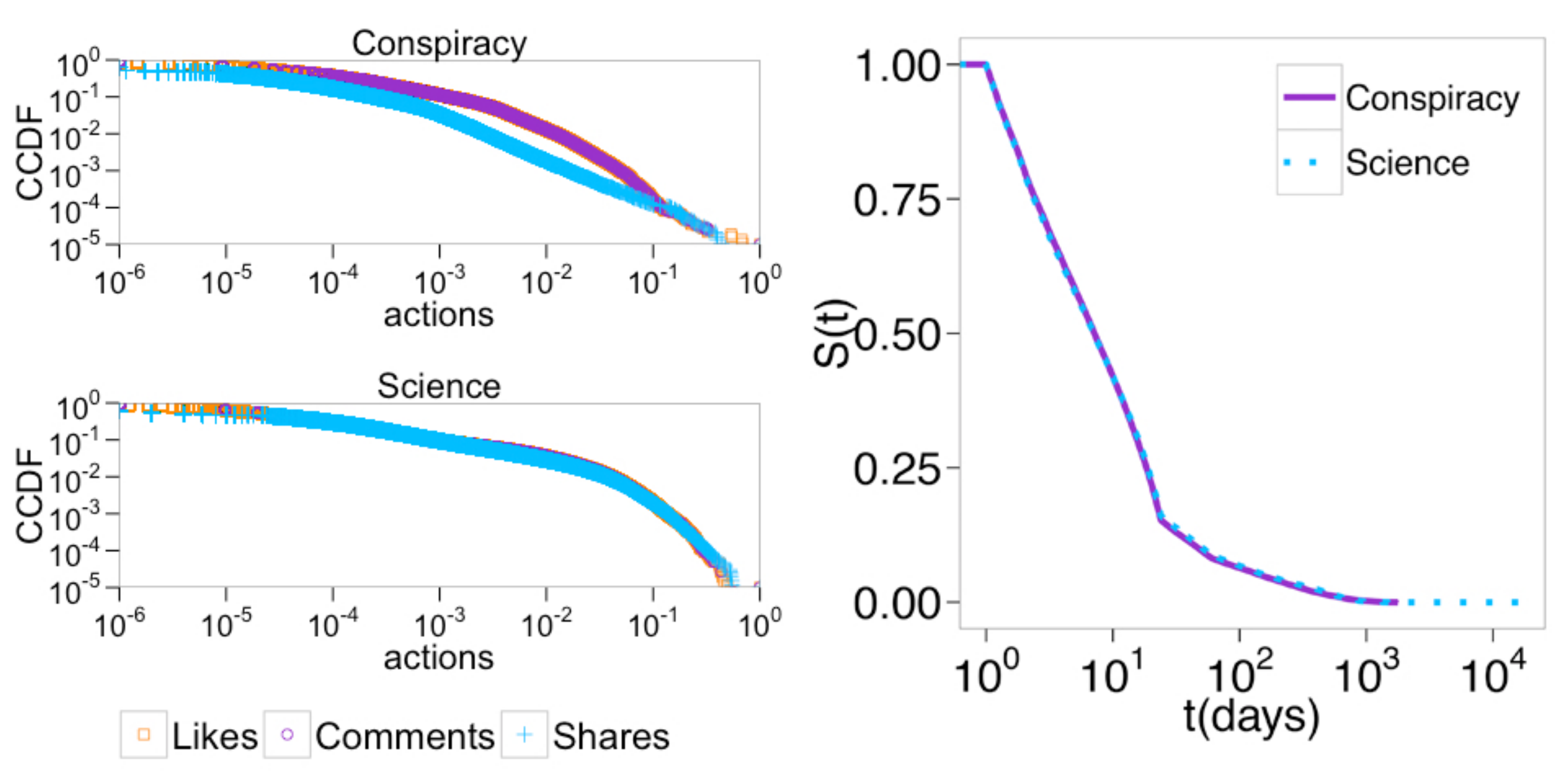} 
	\caption{\textbf{Attention patterns.} \textit{(Left panel)} Complementary cumulative distribution functions (CCDFs) of the number of likes, comments, and shares received by posts belonging to conspiracy \textit{(top)} and scientific \textit{(bottom)} news. \textbf{Posts lifetime.} \textit{(Right panel)} Kaplan-Meier estimates of survival functions of posts belonging to conspiracy and scientific news.}
	\label{fig:post_attention}
\end{figure}

To further understand how these two segregated communities function, we explore how both communities interact with their preferred type of information.
In the left panel of Figure \ref{fig:post_attention} we show the distributions of the number of likes, comments, and shares on posts belonging to both scientific and conspiracy news. As seen from the plots, all the distributions are heavy-tailed -- i.e, all the distributions are best fitted by power laws and all possess similar scaling parameters (see Methods for further details).

Since comments may be intended as a good approximation for the lifetime of a post, in the right panel of Figure \ref{fig:post_attention} we plot the Kaplan-Meier estimates of survival functions by accounting for the first and last comment to each post grouped by category. 
To further characterize differences between the survival functions, we perform the Peto \& Peto \cite{Peto72asymptoticallyefficient} test to detect whether there is a statistically significant difference between the two survival functions. Since we obtain a p-value of $0.944$, we can state that there are not significant statistical differences between the posts' survival functions on both science and conspiracy news. Error bars in the figure are on the order of the size of the symbols. Thus, users attention is similar in both the science and conspiracy echo chambers.

\begin{figure}[H]
	\centering
	\includegraphics[width=\textwidth]{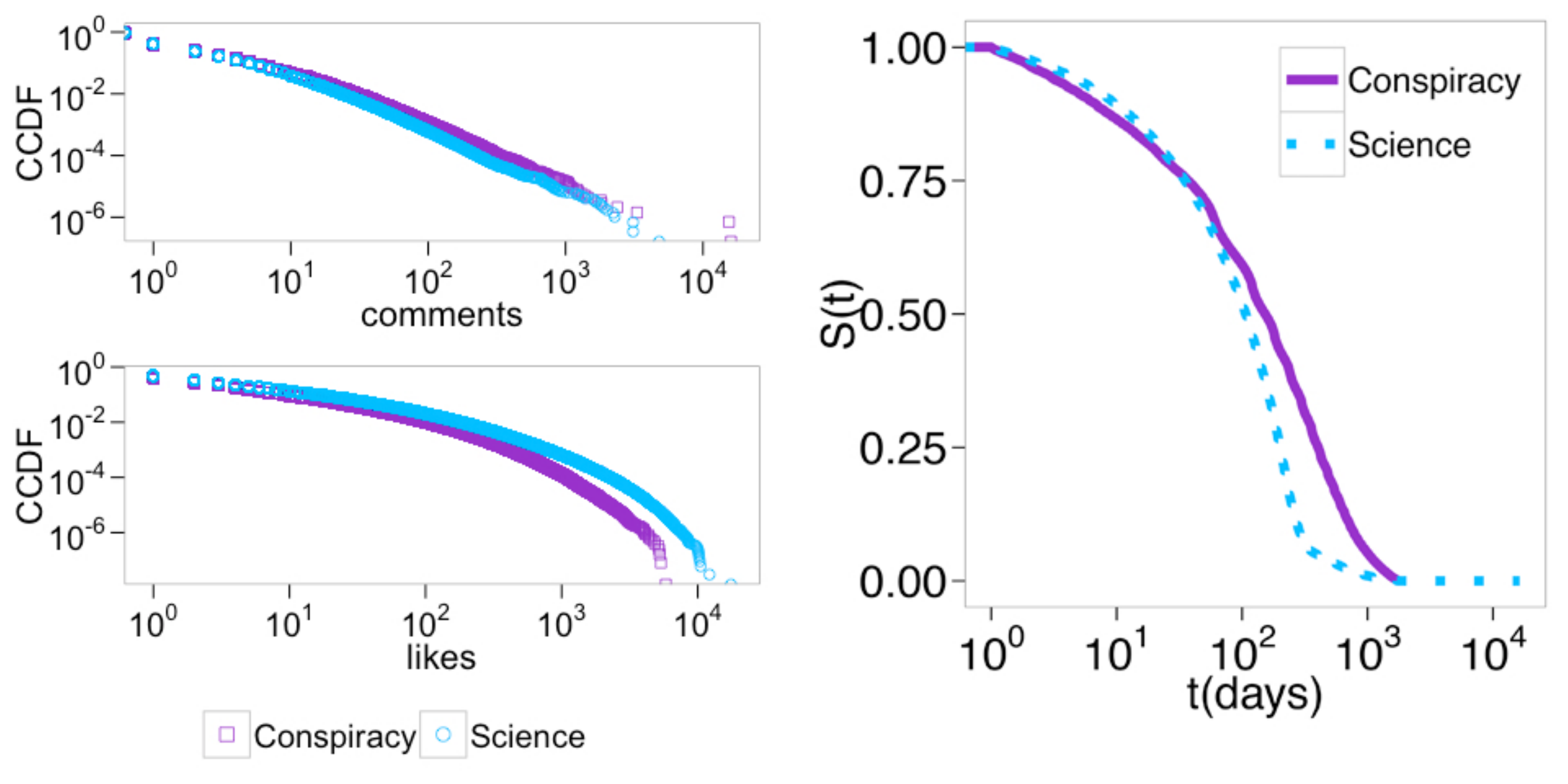} 
	\caption{\textbf{Attention patterns.} (Left panel) Complementary cumulative distribution functions (CCDFs) of the number of comments \textit{(top)}, likes \textit{(bottom)}, per each user on the two categories (conspiracy and science). \textbf{Users lifetime.} (Right panel) Kaplan-Meier estimates of survival functions for users on conspiracy and scientific news.}
	\label{fig:usr_attention}
\end{figure}

We continue our analysis by examining users interaction with different posts on Facebook. In the left panel of Figure \ref{fig:usr_attention} we plot the CCDFs of the number of likes and comments of users on science or conspiracy news. These results show that users consume information in a comparable way -- i.e, \wq{all distributions are heavy tailed} (for scaling parameters and other details refer to Methods section).
\wq{The right panel of Fig. \ref{fig:usr_attention} shows that the persistence of users -- i.e., the Kaplan-Meier estimates of survival functions -- on both types of content is nearly identical. Attention patterns of users in the conspiracy and science echo-chambers reveals that both behave in a very similar manner.}

In summary, contents related to distinct narratives aggregate users into different communities. However, users attention patterns are similar in both communities in terms of interaction and attention to posts. Such a symmetry resembles the idea of echo chambers and filter bubbles -- i.e., users interact only with information that conforms with their system of beliefs and ignore other perspectives and opposing information.

\subsection*{Testing the effect of debunking}
To test the efficacy of debunking attempts and, more generally, to characterize the effect of the exposure to information contrasting a specific narrative, we measure how users -- that we have shown to be aggregated into distinct and polarized communities -- interact with contents aimed at debunking unsubstantiated rumors.  

In a context where users consume and frame narratives based on their immersion in echo chambers, and thus are exposed too just one specific type of content, we hope to understand whether exposure to opposing information is capable of changing user habits and beliefs.
Therefore, we focus on the interaction patterns of polarized conspiracy users (having 95\% of their liking activity on conspiracy rumors) after interacting with debunking posts -- i.e., having commented at least once on a debunking post. 

The first interesting result consists in determining that out of $9,790,906$ polarized conspiracy users, just $117,736$ have interacted with debunking posts --i.e., have commented on a debunking post at least once. \wq{Among these conspiracy users, those with a persistence in the conspiracy echo chamber greater than one day are only $5,831$ --$5,403$ if we consider likes, $2,851$ considering comments. Thus, only a  small fraction of the total conspiracy echo chamber is reached by debunking information.}

\begin{figure}[H]
	\centering
	\includegraphics[width=.7\textwidth]{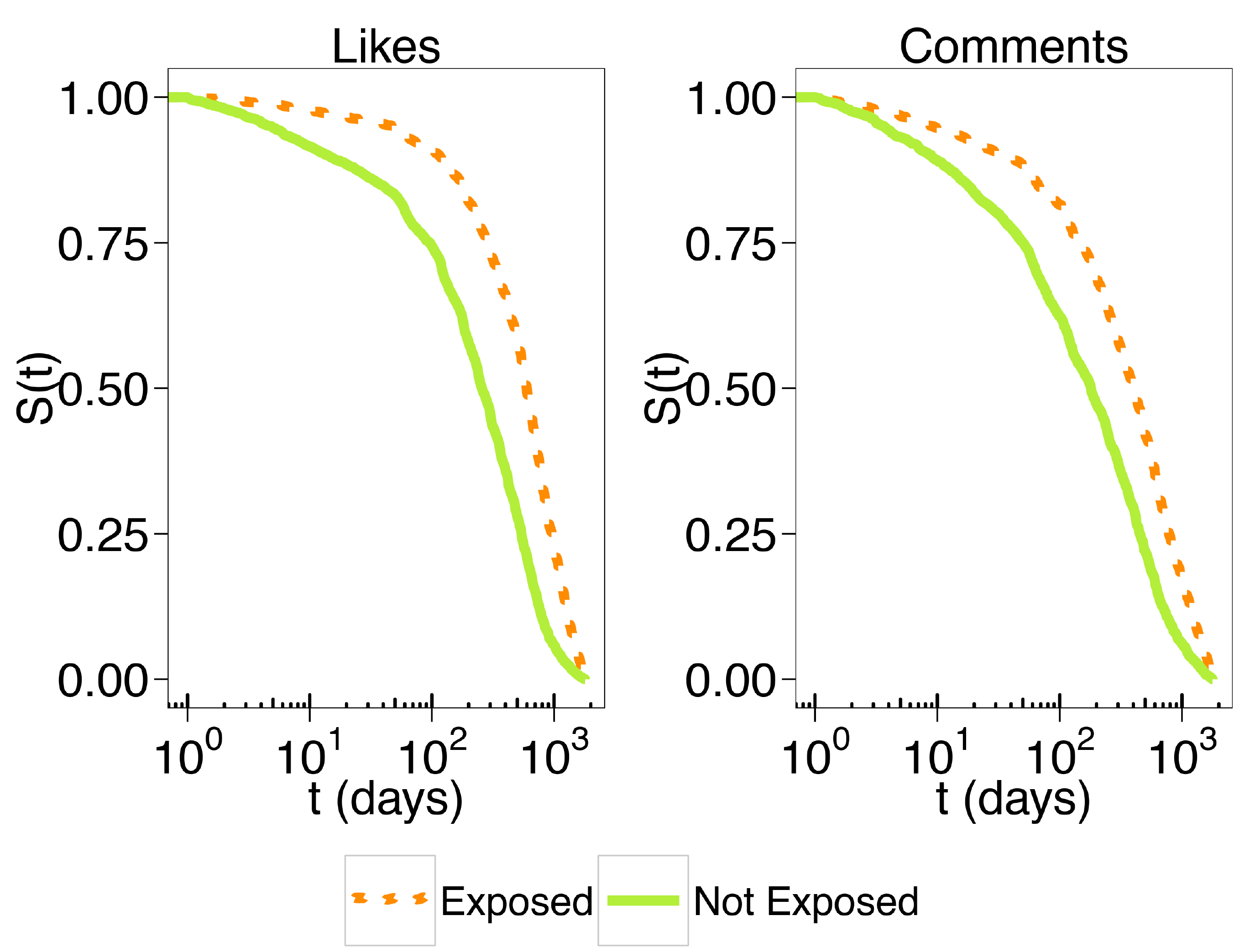} 
	\caption{\textbf{Exposure to debunking.} Kaplan-Meier estimates of survival functions of users exposed and not exposed to debunking. Users lifetime is computed both on their likes \textit{(left)} and comments \textit{(right)}. }
	\label{fig:surv_deb}
\end{figure}

To determine the effect of exposure to debunking posts we analyze the user persistence, defined as the number of days between the first and last like (or comment) on a conspiracy post.  
In Figure \ref{fig:surv_deb} we show the survival functions of conspiracy users exposed and not exposed to debunking posts. 

Table \ref{tab:median} shows the medians of the two distributions and the difference between them, both when the lifetime is computed on likes and on comments.

\begin{table}[ht]
\small
	\centering
	\begin{tabular}{r|c|c}
		& \textbf{Likes} & \textbf{Comments}  \\ \hline
		\textit{Median (Exposed)} & $597.7462$ & $409.8573$ \\ 
		\textit{Median (Not Exposed)} & $249.7239$  & $173.2973$ \\ 
	\end{tabular}
	\caption{\textbf{Medians and difference of Survival functions for lifetime computed both on likes and comments.} }
	\label{tab:median}
\end{table}

\begin{figure}[H]
	\centering
	\includegraphics[width=0.7\textwidth]{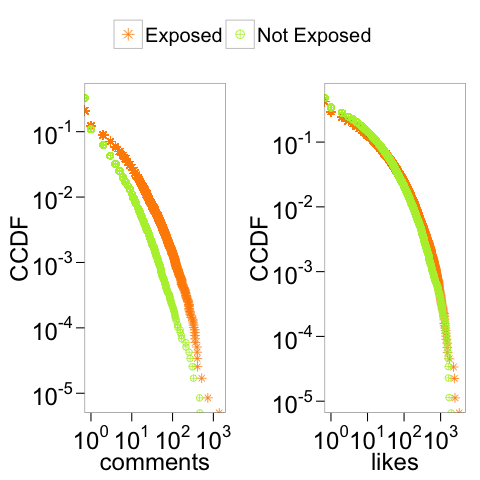} 
	\caption{\textbf{Attention patterns.} Complementary cumulative distribution functions (CCDFs) of the number of comments (top) and likes (bottom), per each user on the two categories (exposed and not exposed to debunking).}
	\label{fig:actions_deb}
\end{figure}

We now compare the liking and commenting patterns of users exposed and not exposed to debunking.
Figure \ref{fig:actions_deb} shows the distributions of the number of comments and likes of the different users. 
\wq{The Spearman's rank correlations coefficient between the number of likes and comments for users exposed and not exposed to debunking are very similar -- $\rho_{exp} = 0.5337494$ (95\%\ c.i. [0.5296521, 0.5378218]); $\rho_{not\_exp} = 0.5698773$ (95\%\ c.i. [0.5660077, 0.5737218]). However, the CCDFs of the number of comments show that users exposed to debunking are in fact slightly \emph{more} prone to comment.}

\begin{figure}[H]
	\centering
	\includegraphics[width=.9\textwidth]{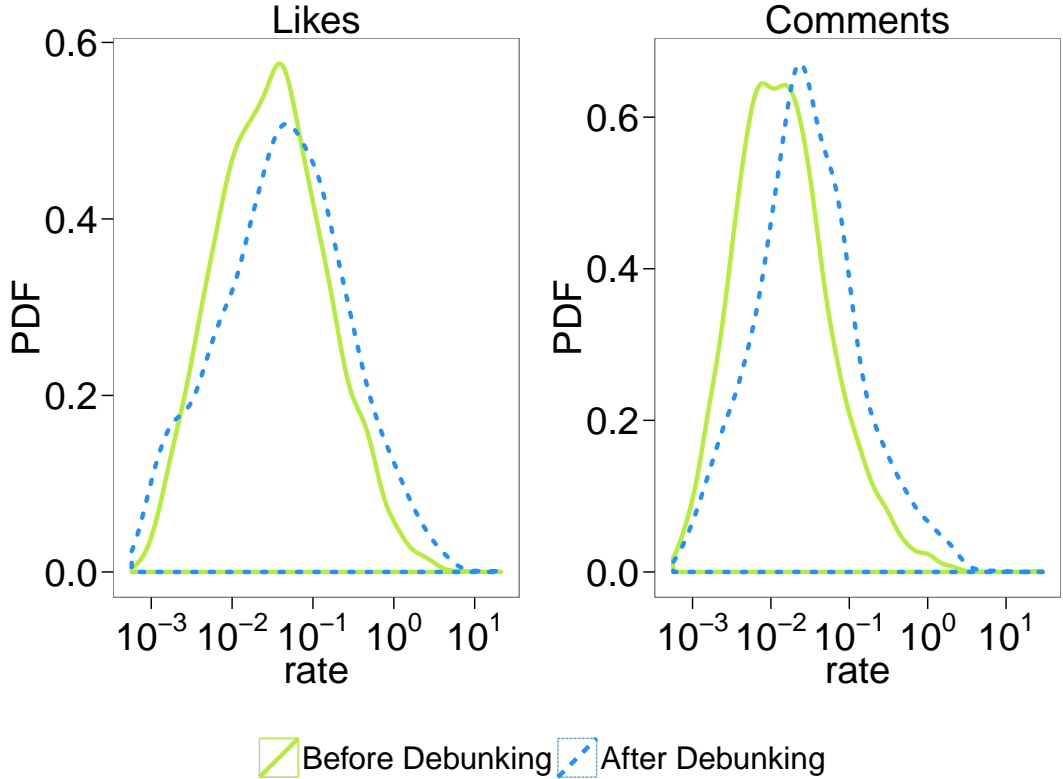} 
	\caption{\textbf{Exposure to debunking.} Rate --i.e.,average number of likes \textit{(left)} (resp., comments \textit{(right)}) on conspiracy posts over time of users exposed to debunking posts.}
	\label{fig:rate_deb}
\end{figure}

At this point we ask whether higher commenting activity is proper of users or it is consequence of the interaction with the debunking post.
To answer this question we analyze users behavior before and after their first interaction (comment) with the debunking post.
Figure \ref{fig:rate_deb} shows the liking and commenting rate -- i.e, the average number of likes (or comments) on conspiracy posts per day -- before and after the first interaction with debunking. 
The plot shows that users' liking and commenting rates increase after the exposure.
To further analyze the effects of interaction with the debunking post we use the Cox Proportional Hazard model \cite{cox} to estimate the hazard of conspiracy users exposed to debunking compared to those not exposed and we find that users not exposed to debunking are $1.76$ times more likely to stop consuming conspiracy news (see Methods section for further details). 

\begin{figure}[H]
	\centering
	\includegraphics[width=\textwidth]{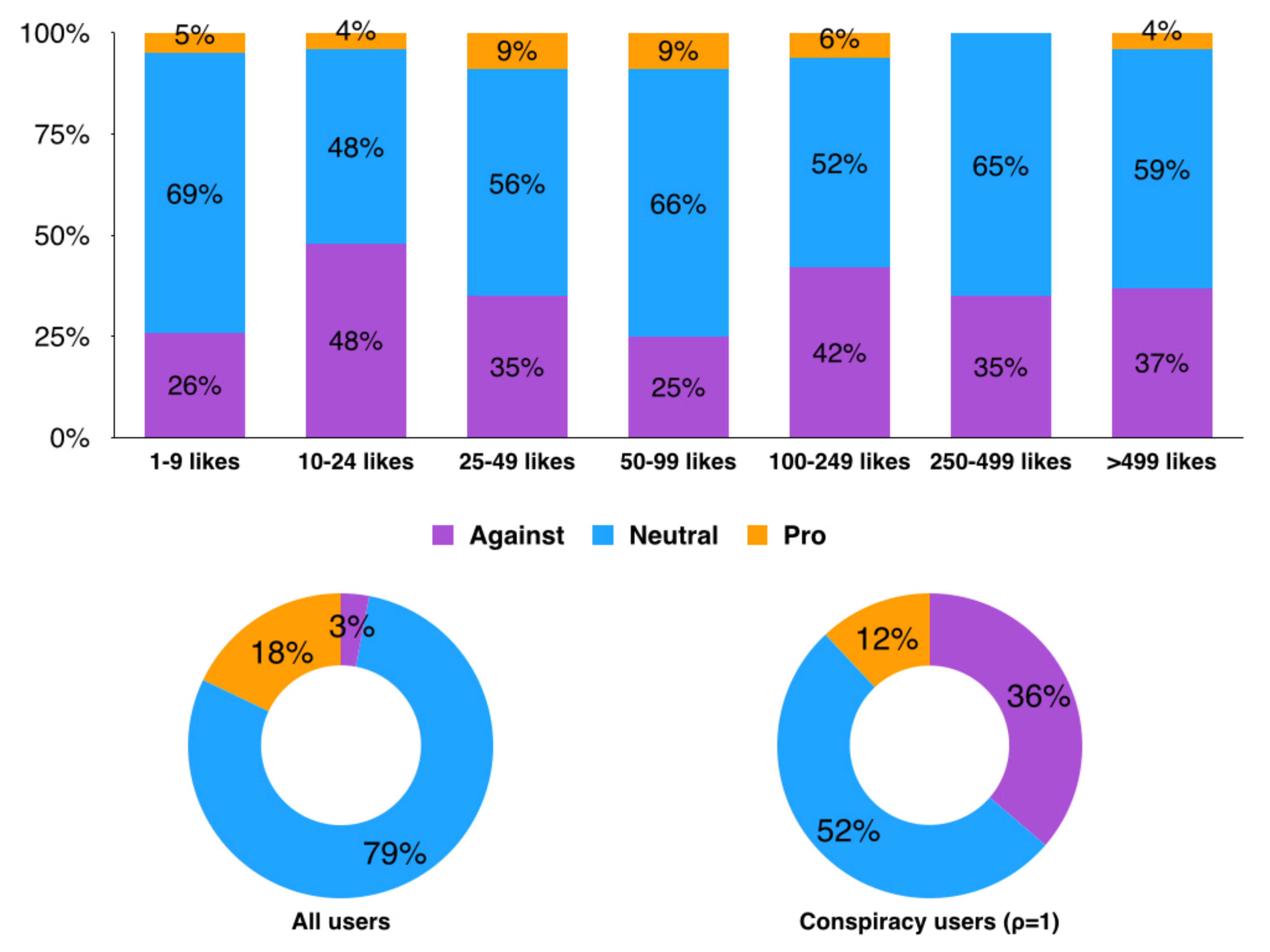} 
	\caption{\textbf{Sentiment on Debunking.} \textit{Top}: Sentiment of comments on debunking posts made by conspiracy users for increasing levels of their number of likes on conspiracy news. \textit{Bottom}: Comparison between the sentiment of all users \textit{(left)} and that of conspiracy users with $\rho=1$ \textit{(right)} on debunking posts. }
	\label{fig:sentiment}
\end{figure}

Finally we analyze the sentiment (\textit{against}, \textit{neutra}l, or \textit{pro}) expressed by conspiracy users when commenting on debunking posts. Our aim is to understand if the level of the engagement in the conspiracy echo chamber constitutes a determinant for the negativity of the sentiment. In the top panel of Figure \ref{fig:sentiment} we show how the sentiment varies when the number of users' likes on conspiracy content increase. Notice that the classification task is intended to tag a comment as negative, neutral, or positive with respect to the post. We randomly selected 100 comments on debunking made by conspiracy users grouped by their engagement in the echo chamber (number of likes on conspiracy posts). Multiple authors classified the posts and we allowed a portion (50\%) of the comments to overlap among all taggers to measure the inter-agreement. By means of a manual inspection iterated over all the authors (with a consensus of 97\%), we confirmed the classifications. We see that comments by conspiracy users are significantly more likely to be negative, regardless of their level of interaction with conspiracies .
Furthermore, in the bottom panel of Figure \ref{fig:sentiment} we compare the behavior of conspiracy users with respect to general users. We randomly select 1000 comments on debunking posts made by all users \textit{(left)} and by conspiracy users having 100\% of their likes on conspiracy posts i.e., $\rho=1$, \textit{(right)}. We notice that users usually exposed to conspiracy like information are significantly more negative when commenting on debunking posts.

In a polarized context where users consume a specific kind of information related to a specific narrative, any attempt to debunk reaches very few users, most of whom react negatively to the undertaken "correction" by increasing their activity on conspiracy pages.

\section*{Conclusions}
Studying the effectiveness of online debunking campaigns is crucial for understanding the processes related to misinformation spreading. \wq{In this work we show the existence of social echo-chambers around different narratives. Two well-formed and highly segregated communities exist around conspiracy and scientific topics -- i.e., users are mainly active in only one category.  Furthermore, by focusing on users interactions with respect to their preferred content, we find similarities in the way that both forms of content are consumed. 
Finally, in order to determine whether online debunking campaigns against false rumors are effective, we measure the response of consumers of conspiracy stories to 47,780 debunking posts.  Our findings shows that  very few users of the conspiracy echo-chamber interact with debunking posts and that, when they do, their interaction often leads to increasing interest in conspiracy-like content. When users are confronted with new and untrusted opposing sources online, the interaction leads them to further commit to their own echo chamber.}

Users tend to aggregate in communities of interests which causes reinforcement and fosters confirmation bias, segregation, polarization, and partisan debates. This comes at the expense of the quality of the information and leads to proliferation of biased narratives fomented by false rumors, mistrust, and paranoia.

Conspiracy related contents become popular because they tend to reduce the complexity of reality and convey general paranoia on specific objects and are more affordable by users. On our perspective the diffusion of bogus content is someway related to the increasing mistrust of people with respect to institutions, to the increasing level of functional illiteracy -- i.e., the inability to understand information correctly-- affecting western countries as well as the combined effect of confirmation bias at work on a enormous basin of information where the quality is poor. 

According to these settings, current debunking campaigns as well as algorithmic solutions do not seem to be the best options. Our findings suggest that the main problem behind misinformation is conservatism rather than gullibility. When users are faced in online discussion with untrusted opponents the discussion results in a major commitment with respect to their own echo chamber.

\section*{Ethical Issues}

The entire data collection process has been carried out exclusively through the Facebook Graph API \cite{fb_graph_api}, which is publicly available, and for the analysis (according to the specification settings of the API) we used only public available data (users with privacy restrictions are not included in the dataset). The pages from which we download data are public Facebook entities (can be accessed by anyone). User content contributing to such pages is also public unless the user's privacy settings specify otherwise and in that case it is not available to us.

\section*{Acknowledgments}
The authors declare no competing interests.
Funding for this work was provided by EU FET project MULTIPLEX nr. 317532, SIMPOL nr. 610704, DOLFINS nr. 640772, SOBIGDATA 654024. The funders had no role in study design, data collection and analysis, decision to publish, or preparation of the manuscript.
We want to thank Geoff All and ``Skepti Forum" for providing fundamental support in defining the atlas of conspiracy news sources in the US Facebook.
Special thanks to Sandro Forgione and Toto Cuturandi for precious suggestions and discussions along the development of this work.

\section*{Methods}
\subsection*{Data Collection}

In this study we address the effect of the usual exposure to diverse verifiable contents to debunking campaingns. 
We identified two main categories of pages: conspiracy news -- i.e. pages promoting contents {\em neglected} by main stream media -- and science news.
Using an approach based on \cite{bessi2014science,mocanu2014}, we defined the space of our investigation with the help of Facebook groups very active in debunking conspiracy theses.
We categorized page according to their contents and their self-description.

Concerning conspiracy news, their self-description is often claiming the mission to inform people about topics neglected by main stream media. Pages as {\em I don't trust the government}, {\em Awakening America}, or {\em Awakened Citizen} promote heterogeneous contents ranging from aliens, chemtrails, geocentrism, up to the causal relation between vaccinations and homosexuality. 
We do not focus on the truth value of their information but rather on the possibility to verify their claims. 
Conversely, science news -- e.g {\em Science}, {\em Science Daily}, {\em Nature} are active in diffusing posts about the most recent scientific advances. 
The selection of the source has been iterated several times and verified by all the authors. 
To our knowledge, the final dataset is the complete set of all scientific and conspiracist information sources active in the US Facebook scenario.
In addition, we identify a set of 66 pages posting debunking information.

The pages from which we downloaded data are public Facebook entities and are accessible by anyone virtually. The entire data collection process is performed exclusively with the Facebook Graph API \cite{fb_graph_api}, which is publicly available and which can be used through one's personal Facebook user account.
The exact breakdown of the data is presented in Table \ref{tab:data_dim}. 
The first category includes all pages diffusing conspiracy information -- pages which disseminate controversial information, most often lacking supporting evidence and sometimes contradictory of the official news (i.e. conspiracy theories). 
The second category is that of scientific dissemination including scientific institutions and scientific press having the main mission to diffuse scientific knowledge. 
The third category contains all pages active in debunking false rumors online. We use this latter set as a testbed for the efficacy of debunking campaign. 

\begin{center}
\begin{table}[!h]
\centering
\begin{tabular}{l|c|c|c|c}
\bf {  }  & \bf {Total} & \bf {Science} & \bf {Conspiracy} & \bf {Debunking} \\ \hline 
Pages & $ 478$ & $ 83$ & $330$ & $66$ \\
Posts & $ 679,948$ & $ 262,815$ & $ 369,420$ & $47,780$ \\
Likes & $603,332,826$ & $453,966,494$ & $145,388,117$  & $3,986,922$ \\
Comments & $30,828,705$ & $22,093,692$ & $8,307,644$ & $429,204$\\
Likers & $52,172,855$ & $39,854,663$ & $19,386,131$& $702,122$\\
Commenters & $9,790,906$ & $7,223,473$ & $3,166,726$ & $118,996$\\
\end{tabular}
  \caption{ \textbf{Breakdown of Facebook dataset.} The number of pages, posts, likes, comments, likers, and commenters for science, conspiracy, and debunking news.}
  \label{tab:data_dim}
\end{table}
\end{center}

\subsection*{Statistical Tools}

To characterize random variables, a main tool is the probability distribution
function (PDF), which gives the probability that a random variable
$X$ assumes a value in the interval $[a,b]$, i.e. $P(a \leq X \leq b) = \int_{a}^{b} f(x) dx$.
The cumulative distribution function (CDF) is another important tool
giving the probability that a random variable $X$ is less than or
equal to a given value $x$, i.e. $F(x) = P(X \leq x) = \int_{-\infty}^{x}f(y)dy$.
In social sciences, an often occuring probability distribution function
is the Pareto's law $f(x) \sim x^{-\gamma}$, that is characterized
by power law tails, i.e. by the occurrence of rare but relevant events.
In fact, while $f(x) \to 0$ for $x \to \infty$ (i.e. high values
of a random variable $X$ are rare), the total probability of rare
events is given by $C(x) = P(X > x) = \int_{x}^{\infty}f(y)dy$, where
$x$ is a sufficiently large value. Notice that $C(x)$ is the Complement
to the CDF (CCDF), where complement indicates that $C(x) = 1 - F(x)$. Hence, in order to better visualize the behavior
of empirical heavy--tailed distributions, we recur to log--log plots
of the CCDF. 

\paragraph{Kaplan-Meier estimator.}
Let us define a random variable $T$ on the interval $[0,\infty)$,
indicating the time an event takes place. The cumulative distribution function (CDF), $F(t) = \textbf{Pr}(T \leq t)$,
indicates the probability that such an event takes place within a given time~$t$. The survival function, defined as the complementary CDF (CCDF\footnote{We remind
that the CCDF of a random variable $X$ is one minus the CDF, the function $f(x)=\textbf{Pr}(X>x)$.}) of $T$, represents the probability that an event lasts beyond a given time period $t$. To estimate this probability we use the \emph{Kaplan--Meier estimator}~\cite{KM58}. 
Let $n_{t}$ denote the number of users who commented before a given time
step $t$, and let $d_{t}$ denote the number of users that stop commenting precisely
at~$t$. Then, the estimated survival probability after time $t$ is
defined as $(n_{t} - d_{t})/n_{t}$. 
Thus, if we have $N$ observations at times $t_1\le t_2\le\dots\le t_N$,
assuming that the events at times $t_i$ are jointly independent, the Kaplan-Meier
estimate of the survival function at time $t$ is defined as
$$\hat{S}(t) = \prod_{t_{i}<t}( \frac{n_{t_i} - d_{t_i}}{n_{t_i}}).$$

\paragraph{Comparison between power law distributions.}
Comparisons between powerlaw distributions of two different quantities -- e.g. likes of different conspiracy topics -- are usually carried out through log-likelihood ratio test \cite{bex2015} or Kolmogorov-Smirnov test \cite{clauset2009}. The former method relies on the ratio between the likelihood of a model fitted on the pooled quantities and the sum of the likelihoods of the models fitted on the two separate quantities, whereas the latter is based on the comparison between the cumulative distribution functions of the two quantities. However, both the afore-mentioned approaches take into account the overall distributions, whereas more often we are especially interested in the scaling parameter of the distribution, i.e. how the tail of the distribution behaves. Moreover, since the Kolmogorov-Smirnov test was conceived for continuous distributions, its application to discrete data gives biased p-values. For these reasons, in this paper we decide to compare our distributions by assess significant differences in the scaling parameters by means of a Wald test. The Wald test we conceive is defined as
$$H_{0} : \hat{\alpha}_{1} - \hat{\alpha}_{2} = 0 $$
$$H_{1} : \hat{\alpha}_{1} - \hat{\alpha}_{2} \neq 0, $$

where $\hat{\alpha}_{1}$ and $\hat{\alpha}_{2}$ are the estimates of the scaling parameters of the two powerlaw distributions. The Wald statistics,
$$  \frac{(\hat{\alpha}_{1} - \hat{\alpha}_{2})^{2}}{VAR(\hat{\alpha}_{1})},$$
where $VAR(\hat{\alpha}_{1})$ is the variance of $\hat{\alpha}_{1}$, follows a $\chi^{2}$ distribution with $1$ degree of freedom. We reject the null hypothesis $H_{0}$ and conclude that there is a significant difference between the scaling parameters of the two distributions if the p-value of the Wald statistics is below a given significance level. 

\subsection*{Attention Patterns}

\subsection*{Content consumption}

Different fits for the tail of the distributions have been taken into account (lognormal, Poisson, exponential, and power law). Goodness of fit tests based on the log-likelihood \cite{clauset2009} have proved that the tails are best fitted by a power law distribution (see Tables \ref{tab:post_gof_con} and \ref{tab:post_gof_sci}). Log-likelihoods of different attention patterns (likes, comments, shares) are computed under competing distributions. The one with the higher log-likelihood is then the better fit \cite{clauset2009}. Log-likelihood ratio tests between power law and the other distributions yield positive ratios, and p-value computed using Vuong's method \cite{vuong} are close to zero, indicating that the best fit provided by the power law distribution is not caused by statistical fluctuations.
Lower bounds and scaling parameters have been estimated via minimization of Kolmogorov-Smirnov statistics \cite{clauset2009}; the latter have been compared via Wald test (see Table \ref{tab:post_est}).

\begin{table}[ht]
	\centering
	\begin{tabular}{r|r|r|r}
		& \textbf{Likes} & \textbf{Comments} & \textbf{Shares} \\ \hline
		\textit{Power law} & $\mathbf{-34,056.95}$ & $\mathbf{-77,904.52}$  & $\mathbf{-108,823.2}$ \\ 
		\textit{Poisson} & $-22,143,084$  & $-6,013,281$   & $-109,045,636$ \\ 
		\textit{Lognormal} & $-35,112.58$ & $-82,619.08$  & $-113,643.7$  \\ 
		\textit{Exponential} & $-36,475.47$ & $-87,859.85$ &  $-119,161.2$  \\ 
	\end{tabular}
	\caption{\textbf{Goodness of fit for attention patterns on Conspiracy pages.} Different fits for the tail of the distributions. Goodness of fit tests based on the log-likelihood proved that the tails are best fitted by a power law distribution}
	\label{tab:post_gof_con}
\end{table}	

\begin{table}[ht]
	\centering
	\begin{tabular}{r|r|r|r}
		& \textbf{Likes} & \textbf{Comments} & \textbf{Shares} \\ \hline
		\textit{Power law} & $\mathbf{-33,371.53}$ & $\mathbf{-2,537.418}$  & $\mathbf{-4,994.981}$ \\ 
		\textit{Poisson} & $-57,731,533$  & $-497,016.2$   & $-3,833,242$ \\ 
		\textit{Lognormal} & $-34,016.76$ & $-2,620.886$  & $-5,126.515$  \\ 
		\textit{Exponential} & $-35.330,76$ & $-2,777.548$ &  $-5,415.722$  \\ 
	\end{tabular}
	\caption{\textbf{Goodness of fit for attention patterns on Science pages.} Different fits for the tail of the distributions. Goodness of fit tests based on the log-likelihood proved that the tails are best fitted by a power law distribution.}
	\label{tab:post_gof_sci}
\end{table}

\begin{table}[ht]
	\centering
	\begin{tabular}{r|cc|cc|cc}
		& \multicolumn{2}{c|}{\textbf{Likes}} & \multicolumn{2}{c|}{\textbf{Comments}} & \multicolumn{2}{c}{\textbf{Shares}}  \\ \hline
		& $\hat{x}_{min}$ & $\hat{\alpha}$ & $\hat{x}_{min}$ & $\hat{\alpha}$ & $\hat{x}_{min}$ & $\hat{\alpha}$ \\ \hline
		\emph{Conspiracy} & $8,995$ & $2.73$ & $136$  & $2.33$ & $1,800$ & $2.29$ \\ 
		\emph{Science}      & $62,976$ & $2.78$ & $8,890$  & $3.27$ & $53,958$ & $3.41$ \\  
		\emph{t-stat} & - & $0.88$ & -& $325.38$ & -& $469.42$ \\
		\emph{p-value} & - & $0.3477$ & -& $< 10^{-6}$ &- & $< 10^{-6}$ \\
		\multicolumn{7}{c}{}\\
	\end{tabular}
	\caption{\textbf{Power law fit of posts' attention patterns.} }
	\label{tab:post_est}
\end{table}

\subsection*{Users Activity}
Table \ref{tab:user_gof_con} and \ref{tab:user_gof_sci} list the fit parameters with various canonical distributions. Table \ref{tab:user_est} shows the power law fit parameters. Table \ref{tab:user_est} summarizes the estimated lower bounds and scaling parameters for each distribution. 

\begin{table}[ht]
	\centering
	\begin{tabular}{r|c|c}
		& \textbf{Likes} & \textbf{Comments}  \\ \hline
		\textit{Power law} & $\mathbf{-24,044.40}$ & $\mathbf{-57,274.31}$ \\ 
		\textit{Poisson} & $-294,076.1$  & $-334,825.6$ \\ 
		\textit{Lognormal} & $-25,177.79$ & $-62,415.91$  \\ 
		\textit{Exponential} & $-28,068.09$ & $-68,650.47$ \\ 
	\end{tabular}
	\caption{\textbf{Goodness of fit for Conspiracy.} }
	\label{tab:user_gof_con}
\end{table}	

\begin{table}[ht]
	\centering
	\begin{tabular}{r|c|c}
		& \textbf{Likes} & \textbf{Comments} \\ \hline
		\textit{Power law} & $\mathbf{-222,763.1}$ & $\mathbf{-42,901.23}$ \\ 
		\textit{Poisson} & $-5,027,337$  & $-260,162.7$  \\ 
		\textit{Lognormal} & $-231,319.1$ & $-46,752.34$   \\ 
		\textit{Exponential} & $-249,771.4$ & $-51,345.45$  \\ 
	\end{tabular}
	\caption{\textbf{Goodness of fit for Science.} }
	\label{tab:user_gof_sci}
\end{table}

\begin{table}[ht]
	\centering
	\begin{tabular}{r|cc|cc}
		& \multicolumn{2}{c|}{\textbf{Likes}} & \multicolumn{2}{c}{\textbf{Comments}}  \\ \hline
		& $\hat{x}_{min}$ & $\hat{\alpha}$ & $\hat{x}_{min}$ & $\hat{\alpha}$\\ \hline
		\emph{Conspiracy} & $900$ & $4.07$ & $45$  & $2.93$  \\ 
		\emph{Science}      & $900$ & $3.25$ & $45$  & $3.07$ \\  
		\emph{t-stat} & & $952.56$ & & $17.89$  \\
		\emph{p-value} & & $< 10^{-6}$ & & $2.34 \times 10^{-5}$ \\
		\multicolumn{5}{c}{}\\
	\end{tabular}
	\caption{\textbf{Power law fit of users' attention patterns.} }
	\label{tab:user_est}
\end{table}

\subsection*{Testing the effect of debunking}

\begin{figure}[H]
	\centering
	\includegraphics[width= .5\textwidth, angle = 0]{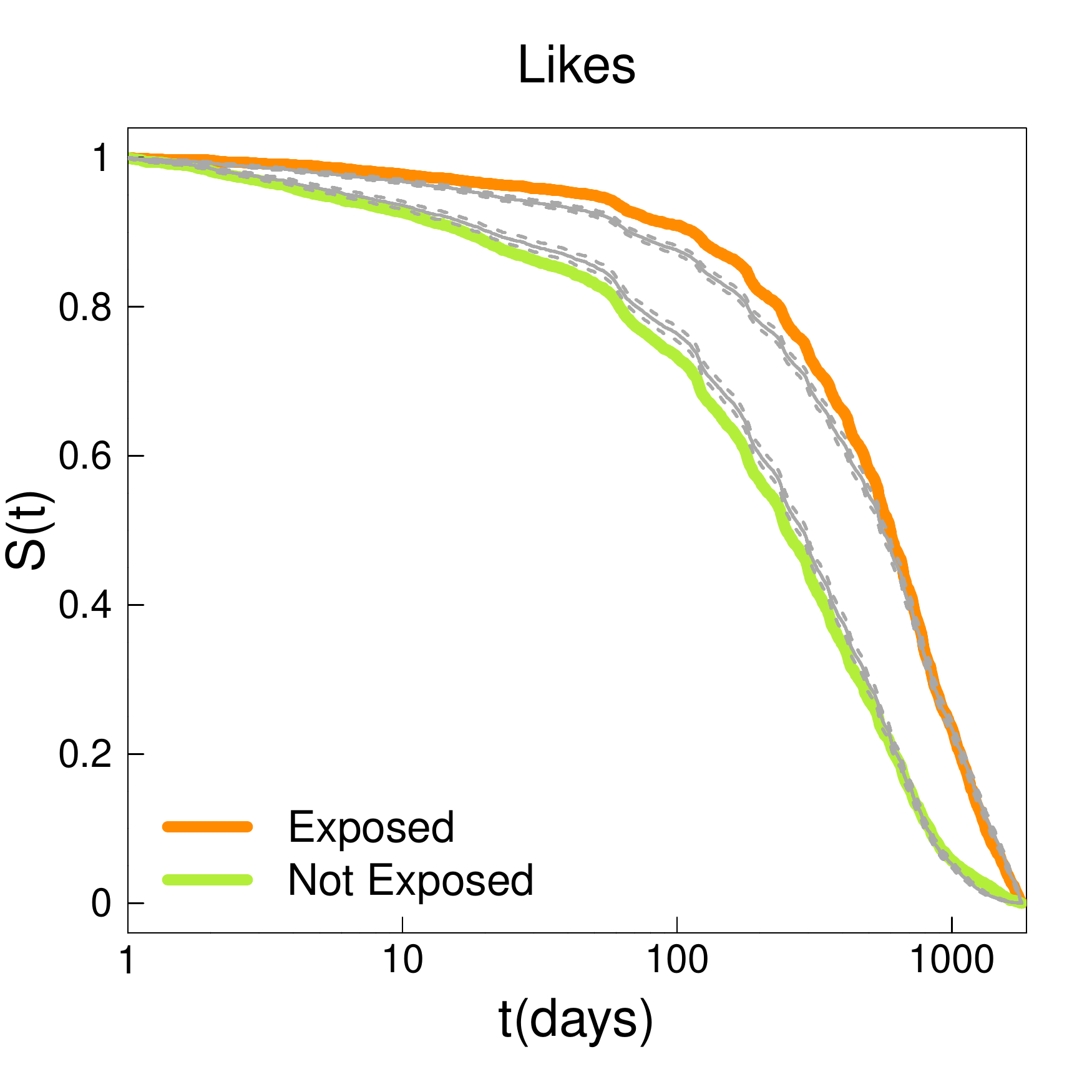} 
	\caption{\textbf{COX-HAZARD MODEL. Likes.} Kaplan-Meier estimates of survival functions of users exposed \textit{(orange)} and not exposed \textit{(green)} to debunking and fits of the Cox proportional hazard model. Lifetime of users is computed on likes.}
	\label{fig:cox_like}
\end{figure}

\begin{figure}[H]
	\centering
	\includegraphics[width=.5\textwidth, angle = 0]{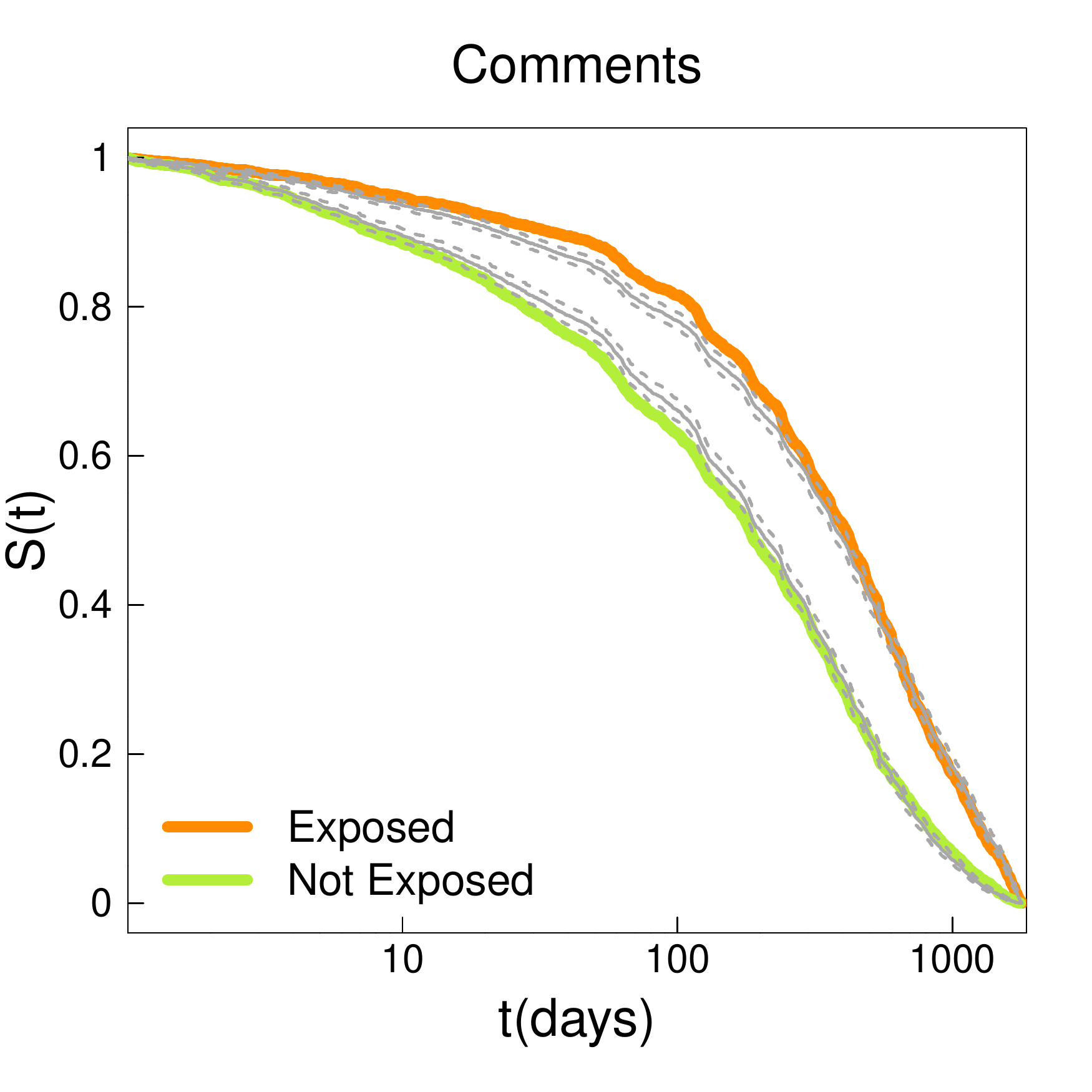} 
	\caption{\textbf{COX-HAZARD MODEL. Comments.}  Kaplan-Meier estimates of survival functions of users exposed \textit{(orange)} and not exposed \textit{(green)} to debunking and fits of the Cox proportional hazard model. Lifetime of users is computed on comments. }
	\label{fig:cox_comm}
\end{figure}

The hazard function is modeled as $h(t) = h_{0}(t) \exp (\beta x)$, where $h_{0}(t)$ is the baseline hazard and $x$ is a dummy variable that takes value $1$ when the user has been exposed to debunking and $0$ otherwise. The hazards depend multiplicatively on the covariates, and $\exp(\beta)$ is the ratio of the hazards between users exposed and not exposed to debunking. The ratio of the hazards of any two users $i$ and $j$ is $\exp(\beta(x_{i}-x_{j}))$, and is called the \emph{hazard ratio}. This ratio is assumed to be constant over time, hence the name of proportional hazard.
When we consider exposure to debunking by means of likes, the estimated $\beta$ is $0.72742\,\, (s.e. = 0.01991,\,\, p < 10^{-6})$ and the corresponding hazard ratio, $\exp(\beta)$, between users exposed and not exposed is $2.07$, indicating that users not exposed to debunking are $2.07$ times more likely to stop consuming conspiracy news. Goodness of fit for the Cox Proportional Hazard Model has been assessed by means of Likelihood ratio test, Wald test, and Score test which provided p-values close to zero. Figure \ref{fig:cox_like} shows the fit of the Cox proportional hazard model.
Moreover, if we consider exposure to debunking by means of comments, the estimated $\beta$ is $0.56748\,\, (s.e. = 0.02711,\,\, p < 10^{-6})$ and the corresponding hazard ratio, $\exp(\beta)$, between users exposed and not exposed is $1.76$, indicating that users not exposed to debunking are $1.76$ times more likely to stop consuming conspiracy news. Goodness of fit for the Cox Proportional Hazard Model has been assessed by means of Likelihood ratio test, Wald test, and Score test which provided p-values close to zero. Figure \ref{fig:cox_comm} shows the fit of the Cox proportional hazard model.

\section*{Pagelists}
\subsection*{Conspiracy pages}

\begin{footnotesize}
\begin{longtable}{|l|l|l|}
  \hline
 & \textbf{Page name} &  \textbf{Page Link}\\ 
  \hline
1 & Spirit Science and Metaphysics & www.facebook.com/171274739679432 \\ 
  2 & Spirit Science & www.facebook.com/210238862349944 \\ 
  3 & The  Conspiracy  Archives & www.facebook.com/262849270399655 \\ 
  4 & iReleaseEndorphins & www.facebook.com/297719273575542 \\ 
  5 & World of Lucid Dreaming & www.facebook.com/98584674825 \\ 
  6 & The Science of Spirit & www.facebook.com/345684712212932 \\ 
  7 & Esoteric Philosophy & www.facebook.com/141347145919527 \\ 
  8 & 9/11 Truth Movement & www.facebook.com/259930617384687 \\ 
  9 & Great Health The Natural Way & www.facebook.com/177320665694370 \\ 
  10 & New World Order News & www.facebook.com/111156025645268 \\ 
  11 & Freedom Isn't Free on FB & www.facebook.com/634692139880441 \\ 
  12 & Skeptic Society & www.facebook.com/224391964369022 \\ 
  13 & The Spiritualist & www.facebook.com/197053767098051 \\ 
  14 & Anonymous World Wide & www.facebook.com/494931210527903 \\ 
  15 & The Life Beyond Earth & www.facebook.com/152806824765696 \\ 
  16 & Illuminati Exposed & www.facebook.com/298088266957281 \\ 
  17 & Illuminating Souls & www.facebook.com/38466722555 \\ 
  18 & Alternative Way & www.facebook.com/119695318182956 \\ 
  19 & Paranormal Conspiracies & www.facebook.com/455572884515474 \\ 
  20 & CANNABIS CURES CANCERS! & www.facebook.com/115759665126597 \\ 
  21 & Natural Cures Not Medicine & www.facebook.com/1104995126306864 \\ 
  22 & CTA Conspiracy Theorists' Association & www.facebook.com/515416211855967 \\ 
  23 & Illuminati Killers & www.facebook.com/478715722175123 \\ 
  24 & Conspiracy 2012 \& Beyond & www.facebook.com/116676015097888 \\ 
  25 & GMO Dangers & www.facebook.com/182443691771352 \\ 
  26 & The Truthers Awareness & www.facebook.com/576279865724651 \\ 
  27 & Exposing the truth about America & www.facebook.com/385979414829070 \\ 
  28 & Occupy Bilderberg & www.facebook.com/231170273608124 \\ 
  29 & Speak the Revolution & www.facebook.com/422518854486140 \\ 
  30 & I Don't Trust The Government & www.facebook.com/380911408658563 \\ 
  31 & Sky Watch Map & www.facebook.com/417198734990619 \\ 
  32 & $|$ truthaholics & www.facebook.com/201546203216539 \\ 
  33 & UFO Phenomenon & www.facebook.com/419069998168962 \\ 
  34 & Conspiracy Theories \& The Illuminati & www.facebook.com/117611941738491 \\ 
  35 & Lets Change The World & www.facebook.com/625843777452057 \\ 
  36 & Makaveli The Prince Killuminati & www.facebook.com/827000284010733 \\ 
  37 & It's A New Day & www.facebook.com/116492031738006 \\ 
  38 & New world outlawz - killuminati soldiers & www.facebook.com/422048874529740 \\ 
  39 & The Government's bullshit. Your argument is invalid. & www.facebook.com/173884216111509 \\ 
  40 & America Awakened & www.facebook.com/620954014584248 \\ 
  41 & The truth behold & www.facebook.com/466578896732948 \\ 
  42 & Alien Ufo And News & www.facebook.com/334372653327841 \\ 
  43 & Anti-Bilderberg Resistance Movement & www.facebook.com/161284443959494 \\ 
  44 & The Truth Unleashed & www.facebook.com/431558836898020 \\ 
  45 & Anti GMO Foods and Fluoride Water & www.facebook.com/366658260094302 \\ 
  46 & STOP Controlling Nature & www.facebook.com/168168276654316 \\ 
  47 & 9/11 Blogger & www.facebook.com/109918092364301 \\ 
  48 & 9/11 Studies and Outreach Club at ASU & www.facebook.com/507983502576368 \\ 
  49 & 9/11 Truth News & www.facebook.com/120603014657906 \\ 
  50 & Abolish the FDA & www.facebook.com/198124706875206 \\ 
  51 & AboveTopSecret.com & www.facebook.com/141621602544762 \\ 
  52 & Activist Post & www.facebook.com/128407570539436 \\ 
  53 & Alliance for Natural Health USA & www.facebook.com/243777274534 \\ 
  54 & All Natural \& Organic. Say No To Toxic Chemicals. & www.facebook.com/323383287739269 \\ 
  55 & Alternative Medicine & www.facebook.com/219403238093061 \\ 
  56 & Alternative World News Network & www.facebook.com/154779684564904 \\ 
  57 & AltHealthWORKS & www.facebook.com/318639724882355 \\ 
  58 & American Academy of Environmental Medicine & www.facebook.com/61115567111 \\ 
  59 & American Association of Naturopathic Physicians & www.facebook.com/14848224715 \\ 
  60 & Ancient Alien Theory & www.facebook.com/147986808591048 \\ 
  61 & Ancient Aliens & www.facebook.com/100140296694563 \\ 
  62 & Ancient Astronaut Theory & www.facebook.com/73808938369 \\ 
  63 & The Anti-Media & www.facebook.com/156720204453023 \\ 
  64 & Anti Sodium Fluoride Movement & www.facebook.com/143932698972116 \\ 
  65 & Architects \& Engineers for 9/11 Truth & www.facebook.com/59185411268 \\ 
  66 & Association of Accredited Naturopathic Medical Colleges (AANMC) & www.facebook.com/60708531146 \\ 
  67 & Autism Media Channel & www.facebook.com/129733027101435 \\ 
  68 & Babes Against Biotech & www.facebook.com/327002374043204 \\ 
  69 & Bawell Alkaline Water Ionizer Health Benefits & www.facebook.com/447465781968559 \\ 
  70 & CancerTruth & www.facebook.com/348939748204 \\ 
  71 & Chemtrails Awareness & www.facebook.com/12282631069 \\ 
  72 & Collective Evolution & www.facebook.com/131929868907 \\ 
  73 & Conspiracy Theory With Jesse Ventura & www.facebook.com/122021024620821 \\ 
  74 & The Daily Sheeple & www.facebook.com/114637491995485 \\ 
  75 & Dr. Bronner's Magic Soaps & www.facebook.com/33699882778 \\ 
  76 & Dr. Joseph Mercola & www.facebook.com/114205065589 \\ 
  77 & Dr. Ronald Hoffman & www.facebook.com/110231295707464 \\ 
  78 & Earth. We are one. & www.facebook.com/149658285050501 \\ 
  79 & Educate Inspire Change & www.facebook.com/467083626712253 \\ 
  80 & Energise for Life: The Alkaline Diet Experts! & www.facebook.com/99263884780 \\ 
  81 & Exposing The Truth & www.facebook.com/175868780941 \\ 
  82 & The Farmacy & www.facebook.com/482134055140366 \\ 
  83 & Fluoride Action Network & www.facebook.com/109230302473419 \\ 
  84 & Food Babe & www.facebook.com/132535093447877 \\ 
  85 & Global Research (Centre for Research on Globalization) & www.facebook.com/200870816591393 \\ 
  86 & GMO Inside & www.facebook.com/478981558808326 \\ 
  87 & GMO Just Say No & www.facebook.com/1390244744536466 \\ 
  88 & GreenMedInfo.com & www.facebook.com/111877548489 \\ 
  89 & Healthy Holistic Living & www.facebook.com/134953239880777 \\ 
  90 & I Fucking Love Truth & www.facebook.com/445723122122920 \\ 
  91 & InfoWars & www.facebook.com/80256732576 \\ 
  92 & Institute for Responsible Technology & www.facebook.com/355853721234 \\ 
  93 & I Want To Be 100\% Organic & www.facebook.com/431825520263804 \\ 
  94 & Knowledge of Today & www.facebook.com/307551552600363 \\ 
  95 & La Healthy Living & www.facebook.com/251131238330504 \\ 
  96 & March Against Monsanto & www.facebook.com/566004240084767 \\ 
  97 & Millions Against Monsanto by OrganicConsumers.org & www.facebook.com/289934516904 \\ 
  98 & The Mind Unleashed & www.facebook.com/432632306793920 \\ 
  99 & Moms Across America & www.facebook.com/111116155721597 \\ 
  100 & Moms for Clean Air/Stop Jet Aerosol Spraying & www.facebook.com/1550135768532988 \\ 
  101 & Natural Society & www.facebook.com/191822234195749 \\ 
  102 & Non-GMO Project & www.facebook.com/55972693514 \\ 
  103 & Occupy Corporatism & www.facebook.com/227213404014035 \\ 
  104 & The Open Mind & www.facebook.com/782036978473504 \\ 
  105 & Organic Consumers Association & www.facebook.com/13341879933 \\ 
  106 & Organic Health & www.facebook.com/637019016358534 \\ 
  107 & The Organic Prepper & www.facebook.com/435427356522981 \\ 
  108 & PreventDisease.com & www.facebook.com/199701427498 \\ 
  109 & Raw For Beauty & www.facebook.com/280583218719915 \\ 
  110 & REALfarmacy.com & www.facebook.com/457765807639814 \\ 
  111 & ReThink911 & www.facebook.com/581078305246370 \\ 
  112 & Sacred Geometry and Ancient Knowledge & www.facebook.com/363116270489862 \\ 
  113 & Stop OC Smart Meters & www.facebook.com/164620026961366 \\ 
  114 & The Top Information Post & www.facebook.com/505941169465529 \\ 
  115 & The Truth About Vaccines & www.facebook.com/133579170019140 \\ 
  116 & Truth Teller & www.facebook.com/278837732170258 \\ 
  117 & Veterans Today & www.facebook.com/170917822620 \\ 
  118 & What Doctors Don't Tell You & www.facebook.com/157620297591924 \\ 
  119 & Wheat Belly & www.facebook.com/209766919069873 \\ 
  120 & Why don't you try this? & www.facebook.com/202719226544269 \\ 
  121 & WND & www.facebook.com/119984188013847 \\ 
  122 & WorldTruth.TV & www.facebook.com/114896831960040 \\ 
  123 & Zeitgeist & www.facebook.com/32985985640 \\ 
  124 & Ancient Origins & www.facebook.com/530869733620642 \\ 
  125 & Astrology Answers & www.facebook.com/413145432131383 \\ 
  126 & Astrology News Service & www.facebook.com/196416677051124 \\ 
  127 & Autism Action Network & www.facebook.com/162315170489749 \\ 
  128 & Awakening America & www.facebook.com/406363186091465 \\ 
  129 & Awakening People & www.facebook.com/204136819599624 \\ 
  130 & Cannabinoids Cure Diseases \& The Endocannabinoid System Makes It Possible. & www.facebook.com/322971327723145 \\ 
  131 & Celestial Healing Wellness Center & www.facebook.com/123165847709982 \\ 
  132 & Chico Sky Watch & www.facebook.com/149772398420200 \\ 
  133 & A Conscious awakening & www.facebook.com/539906446080416 \\ 
  134 & Conspiracy Syndrome & www.facebook.com/138267619575029 \\ 
  135 & Conspiracy Theory: Truth Hidden in Plain Sight, and Army of SATAN & www.facebook.com/124113537743088 \\ 
  136 & Cosmic Intelligence-Agency & www.facebook.com/164324963624932 \\ 
  137 & C4ST & www.facebook.com/371347602949295 \\ 
  138 & Deepak Chopra & www.facebook.com/184133190664 \\ 
  139 & Dr. Mehmet Oz & www.facebook.com/35541499994 \\ 
  140 & Earth Patriot & www.facebook.com/373323356902 \\ 
  141 & Electromagnetic Radiation Safety & www.facebook.com/465980443450930 \\ 
  142 & EMF Safety Network & www.facebook.com/199793306742863 \\ 
  143 & End Time Headlines & www.facebook.com/135010313189665 \\ 
  144 & Young Living Essential Oils & www.facebook.com/29796911981 \\ 
  145 & Exposing Bilderberg 2012 & www.facebook.com/300498383360728 \\ 
  146 & Exposing The Illuminati & www.facebook.com/196087297165394 \\ 
  147 & Exposing Satanic World Government & www.facebook.com/529736240478567 \\ 
  148 & FEMA Camps Exposed & www.facebook.com/285257418255898 \\ 
  149 & Fight Against Illuminati And New World Order & www.facebook.com/195559810501401 \\ 
  150 & FitLife.tv & www.facebook.com/148518475178805 \\ 
  151 & GMO Free USA & www.facebook.com/402058139834655 \\ 
  152 & Holistic Health & www.facebook.com/105497186147476 \\ 
  153 & The Illuminati & www.facebook.com/543854275628660 \\ 
  154 & Illuminati Mind Control & www.facebook.com/499866223357022 \\ 
  155 & Intelwars & www.facebook.com/130166550361356 \\ 
  156 & Natural Solutions Foundation & www.facebook.com/234136166735798 \\ 
  157 & NWO Truth Radio & www.facebook.com/135090269995781 \\ 
  158 & Occupy Bilderberg 2012 & www.facebook.com/227692450670795 \\ 
  159 & Operation: Awakening- The Global Revolution & www.facebook.com/287772794657070 \\ 
  160 & The Paradigm Shift & www.facebook.com/221341527884801 \\ 
  161 & PositiveMed & www.facebook.com/177648308949017 \\ 
  162 & Press TV & www.facebook.com/145097112198751 \\ 
  163 & The Resistance & www.facebook.com/394604877344757 \\ 
  164 & Rima E. Laibow, M.D. - Save My Life Dr. Rima & www.facebook.com/107527312740569 \\ 
  165 & RT America & www.facebook.com/137767151365 \\ 
  166 & Ruble's Wonderings - Forbidden Archeology \& Science & www.facebook.com/265422293590870 \\ 
  167 & Seekers Of Truth & www.facebook.com/736499966368634 \\ 
  168 & Spiritual Ecology & www.facebook.com/261982733906722 \\ 
  169 & Spiritualer.com & www.facebook.com/531950866874307 \\ 
  170 & Take Back Your Power & www.facebook.com/269179579827247 \\ 
  171 & There is a cure for Cancer, but it is not FDA approved. Phoenix Tears work! & www.facebook.com/395190597537 \\ 
  172 & True Activist & www.facebook.com/129370207168068 \\ 
  173 & Truth Exposed Radio & www.facebook.com/173823575962481 \\ 
  174 & Truth Movement & www.facebook.com/161389033958012 \\ 
  175 & Truth Network & www.facebook.com/271701606246002 \\ 
  176 & Wake up call & www.facebook.com/276404442375280 \\ 
  177 & We Should Ban GMOs & www.facebook.com/516524895097781 \\ 
  178 & vactruth.com & www.facebook.com/287991907988 \\ 
  179 & Veterans Today Truth Warriors & www.facebook.com/645478795537771 \\ 
  180 & 4 Foot Farm Blueprint & www.facebook.com/1377091479178258 \\ 
  181 & Dawning Golden Crystal Age & www.facebook.com/127815003927694 \\ 
  182 & Occupy Your Mind & www.facebook.com/393849780700637 \\ 
  183 & We do not Forgive. We do not Forget. We are Anonymous. Expect Us. & www.facebook.com/134030470016833 \\ 
  184 & Health Impact News & www.facebook.com/469121526459635 \\ 
  185 & NaturalNews.com & www.facebook.com/35590531315 \\ 
  186 & World for 9/11 Truth & www.facebook.com/38411749990 \\ 
  187 & Beware of Disinformation & www.facebook.com/558882824140805 \\ 
  188 & Citizens For Legitimate Government & www.facebook.com/93486533659 \\ 
  189 & Cureyourowncancer.org & www.facebook.com/535679936458252 \\ 
  190 & Juicing Vegetables & www.facebook.com/172567162798498 \\ 
  191 & Quantum Prophecies & www.facebook.com/323520924404870 \\ 
  192 & AIM Integrative Medicine & www.facebook.com/137141869763519 \\ 
  193 & Autism Nutrition Research Center & www.facebook.com/1508552969368252 \\ 
  194 & The Canary Party & www.facebook.com/220071664686886 \\ 
  195 & Chemtrail Research & www.facebook.com/247681531931261 \\ 
  196 & Chemtrail Watchers & www.facebook.com/77065926441 \\ 
  197 & Children's Medical Safety Research Institute & www.facebook.com/790296257666848 \\ 
  198 & Contaminated Vaccines & www.facebook.com/686182981422650 \\ 
  199 & Dane Wigington & www.facebook.com/680418385353616 \\ 
  200 & David Icke & www.facebook.com/147823328841 \\ 
  201 & David Icke Books Limited & www.facebook.com/191364871070270 \\ 
  202 & David Icke - Headlines & www.facebook.com/1421025651509652 \\ 
  203 & Disinformation Directory & www.facebook.com/258624097663749 \\ 
  204 & The Drs. Wolfson & www.facebook.com/1428115297409777 \\ 
  205 & Educate, Inspire \& Change.  The Truth Is Out There, Just Open Your Eyes & www.facebook.com/111415972358133 \\ 
  206 & Focus for Health Foundation & www.facebook.com/456051981200997 \\ 
  207 & Generation Rescue & www.facebook.com/162566388038 \\ 
  208 & Geoengineering Watch & www.facebook.com/448281071877305 \\ 
  209 & Global Skywatch & www.facebook.com/128141750715760 \\ 
  210 & The Greater Good & www.facebook.com/145865008809119 \\ 
  211 & The Health Freedom Express & www.facebook.com/450411098403289 \\ 
  212 & Homegrown Health & www.facebook.com/190048467776279 \\ 
  213 & Intellihub & www.facebook.com/439119036166643 \\ 
  214 & The Liberty Beacon & www.facebook.com/222092971257181 \\ 
  215 & International Medical Council on Vaccination & www.facebook.com/121591387888250 \\ 
  216 & International Medical Council on Vaccination - Maine Chapter & www.facebook.com/149150225097217 \\ 
  217 & Medical Jane & www.facebook.com/156904131109730 \\ 
  218 & Mississippi Parents for Vaccine Rights & www.facebook.com/141170989357307 \\ 
  219 & My parents didn't put me in time-out, they whooped my ass! & www.facebook.com/275738084532 \\ 
  220 & National Vaccine Information Center & www.facebook.com/143745137930 \\ 
  221 & The Raw Feed Live & www.facebook.com/441287025913792 \\ 
  222 & Rinf.com & www.facebook.com/154434341237962 \\ 
  223 & SANEVAX & www.facebook.com/139881632707155 \\ 
  224 & Things pro-vaxers say & www.facebook.com/770620782980490 \\ 
  225 & Unvaccinated America & www.facebook.com/384030984975351 \\ 
  226 & Vaccine Injury Law Project & www.facebook.com/295977950440133 \\ 
  227 & Vermont Coalition for Vaccine Choice & www.facebook.com/380959335251497 \\ 
  228 & 9/11: The BIGGEST LIE & www.facebook.com/129496843915554 \\ 
  229 & Agent Orange Activists & www.facebook.com/644062532320637 \\ 
  230 & Age of Autism & www.facebook.com/183383325034032 \\ 
  231 & AutismOne & www.facebook.com/199957646696501 \\ 
  232 & Awakened Citizen & www.facebook.com/481936318539426 \\ 
  233 & Best Chinese Medicines & www.facebook.com/153901834710826 \\ 
  234 & Black Salve & www.facebook.com/224002417695782 \\ 
  235 & Bought Movie & www.facebook.com/144198595771434 \\ 
  236 & Children Of Vietnam Veterans Health Alliance & www.facebook.com/222449644516926 \\ 
  237 & Collective-Evolution Shift & www.facebook.com/277160669144420 \\ 
  238 & Doctors Are Dangerous & www.facebook.com/292077004229528 \\ 
  239 & Dr. Tenpenny on Vaccines & www.facebook.com/171964245890 \\ 
  240 & Dr Wakefield's work must continue & www.facebook.com/84956903164 \\ 
  241 & EndoRIOT & www.facebook.com/168746323267370 \\ 
  242 & Enenews & www.facebook.com/126572280756448 \\ 
  243 & Expanded Consciousness & www.facebook.com/372843136091545 \\ 
  244 & Exposing the truths of the Illuminati II & www.facebook.com/157896884221277 \\ 
  245 & Family Health Freedom Network & www.facebook.com/157276081149274 \\ 
  246 & Fearless Parent & www.facebook.com/327609184049041 \\ 
  247 & Food Integrity Now & www.facebook.com/336641393949 \\ 
  248 & Four Winds 10 & www.facebook.com/233310423466959 \\ 
  249 & Fukushima Explosion What You Do Not Know & www.facebook.com/1448402432051510 \\ 
  250 & The Golden Secrets & www.facebook.com/250112083847 \\ 
  251 & Health Without Medicine \& Food Without Chemicals & www.facebook.com/304937512905083 \\ 
  252 & Higher Perspective & www.facebook.com/488353241197000 \\ 
  253 & livingmaxwell & www.facebook.com/109584749954 \\ 
  254 & JFK Truth & www.facebook.com/1426437510917392 \\ 
  255 & New World Order Library $|$ NWO Library & www.facebook.com/194994541179 \\ 
  256 & No Fluoride & www.facebook.com/117837414684 \\ 
  257 & Open Minds Magazine & www.facebook.com/139382669461984 \\ 
  258 & Organic Seed Alliance & www.facebook.com/111220277149 \\ 
  259 & Organic Seed Growers and Trade Association & www.facebook.com/124679267607065 \\ 
  260 & RadChick Radiation Research \& Mitigation & www.facebook.com/260610960640885 \\ 
  261 & The REAL Institute - Max Bliss & www.facebook.com/328240720622120 \\ 
  262 & Realities Watch & www.facebook.com/647751428644641 \\ 
  263 & StormCloudsGathering & www.facebook.com/152920038142341 \\ 
  264 & Tenpenny Integrative Medical Centers (TIMC) & www.facebook.com/144578885593545 \\ 
  265 & Vaccine Epidemic & www.facebook.com/190754844273581 \\ 
  266 & VaccineImpact & www.facebook.com/783513531728629 \\ 
  267 & Weston A. Price Foundation & www.facebook.com/58956225915 \\ 
  268 & What On Earth Is Happening & www.facebook.com/735263086566914 \\ 
  269 & The World According to Monsanto & www.facebook.com/70550557294 \\ 
  270 & Truth Theory & www.facebook.com/175719755481 \\ 
  271 & Csglobe & www.facebook.com/403588786403016 \\ 
  272 & Free Energy Truth & www.facebook.com/192446108025 \\ 
  273 & Smart Meter Education Network & www.facebook.com/630418936987737 \\ 
  274 & The Mountain Astrologer magazine & www.facebook.com/112278112664 \\ 
  275 & Alberta Chemtrail Crusaders & www.facebook.com/1453419071541217 \\ 
  276 & Alkaline Us & www.facebook.com/430099307105773 \\ 
  277 & Americas Freedom Fighters & www.facebook.com/568982666502934 \\ 
  278 & Anti-Masonic Party Founded 1828 & www.facebook.com/610426282420191 \\ 
  279 & Cannabidiol OIL & www.facebook.com/241449942632203 \\ 
  280 & Cancer Compass\~{}An Alternate Route & www.facebook.com/464410856902927 \\ 
  281 & Collective Evolution Lifestyle & www.facebook.com/1412660665693795 \\ 
  282 & Conscious Life News & www.facebook.com/148270801883880 \\ 
  283 & Disclosure Project & www.facebook.com/112617022158085 \\ 
  284 & Dr. Russell Blaylock, MD & www.facebook.com/123113281055091 \\ 
  285 & Dumbing Down People into Sheeple & www.facebook.com/123846131099156 \\ 
  286 & Expand Your Consciousness & www.facebook.com/351484988331613 \\ 
  287 & Fluoride: Poison on Tap & www.facebook.com/1391282847818928 \\ 
  288 & Gaiam TV & www.facebook.com/182073298490036 \\ 
  289 & Gary Null \& Associates & www.facebook.com/141821219197583 \\ 
  290 & Genesis II Church of Health \& Healing (Official) & www.facebook.com/115744595234934 \\ 
  291 & Genetic Crimes Unit & www.facebook.com/286464338091839 \\ 
  292 & Global Healing Center & www.facebook.com/49262013645 \\ 
  293 & Gluten Free Society & www.facebook.com/156656676820 \\ 
  294 & GMO Free Oregon & www.facebook.com/352284908147199 \\ 
  295 & GMO Journal & www.facebook.com/113999915313056 \\ 
  296 & GMO OMG & www.facebook.com/525732617477488 \\ 
  297 & GreenMedTV & www.facebook.com/1441106586124552 \\ 
  298 & Healing The Symptoms Known As Autism & www.facebook.com/475607685847989 \\ 
  299 & Health Conspiracy Radio & www.facebook.com/225749987558859 \\ 
  300 & Health and Happiness & www.facebook.com/463582507091863 \\ 
  301 & Jesse Ventura & www.facebook.com/138233432870955 \\ 
  302 & Jim Humble & www.facebook.com/252310611483446 \\ 
  303 & Kid Against Chemo & www.facebook.com/742946279111241 \\ 
  304 & Kids Right To Know Club & www.facebook.com/622586431101931 \\ 
  305 & The Master Mineral Solution of the 3rd Millennium & www.facebook.com/527697750598681 \\ 
  306 & Millions Against Monsanto Maui & www.facebook.com/278949835538988 \\ 
  307 & Millions Against Monsanto World Food Day 2011 & www.facebook.com/116087401827626 \\ 
  308 & Newsmax Health & www.facebook.com/139852149523097 \\ 
  309 & Non GMO journal & www.facebook.com/303024523153829 \\ 
  310 & Nurses Against ALL Vaccines & www.facebook.com/751472191586573 \\ 
  311 & Oath Keepers & www.facebook.com/182483688451972 \\ 
  312 & Oath Keepers of America & www.facebook.com/1476304325928788 \\ 
  313 & The Organic \& Non-GMO Report & www.facebook.com/98397470347 \\ 
  314 & Oregon Coast Holographic Skies Informants & www.facebook.com/185456364957528 \\ 
  315 & Paranormal Research Project & www.facebook.com/1408287352721685 \\ 
  316 & Politically incorrect America & www.facebook.com/340862132747401 \\ 
  317 & (Pure Energy Systems) PES Network, Inc. & www.facebook.com/183247495049420 \\ 
  318 & Save Hawaii from Monsanto & www.facebook.com/486359274757546 \\ 
  319 & Sayer Ji & www.facebook.com/205672406261058 \\ 
  320 & SecretSpaceProgram & www.facebook.com/126070004103888 \\ 
  321 & SPM  Southern Patriots MIlitia & www.facebook.com/284567008366903 \\ 
  322 & Thrive & www.facebook.com/204987926185574 \\ 
  323 & Truth Connections & www.facebook.com/717024228355607 \\ 
  324 & Truth Frequency & www.facebook.com/396012345346 \\ 
  325 & Truthstream Media.com & www.facebook.com/193175867500745 \\ 
  326 & VT Right To Know GMOs & www.facebook.com/259010264170581 \\ 
  327 & We Are Change & www.facebook.com/86518833689 \\ 
  328 & Wisdom Tribe 7 Walking in Wisdom. & www.facebook.com/625899837467523 \\ 
  329 & World Association for Vaccine Education & www.facebook.com/1485654141655627 \\ 
  330 & X Tribune & www.facebook.com/1516605761946273 \\ 
   \hline
\end{longtable}
\end{footnotesize}

\newpage
\subsection*{Science pages}

\begin{footnotesize}
\begin{longtable}{|l|l|l|}
  \hline
  & \textbf{Page name} &  \textbf{Page Link}\\ 
  \hline
1 & AAAS - The American Association for the Advancement of Science & www.facebook.com/19192438096 \\ 
  2 & AAAS Dialogue on Science, Ethics and Religion & www.facebook.com/183292605082365 \\ 
  3 & Armed with Science & www.facebook.com/228662449288 \\ 
  4 & AsapSCIENCE & www.facebook.com/162558843875154 \\ 
  5 & Bridge to Science & www.facebook.com/185160951530768 \\ 
  6 & EurekAlert! & www.facebook.com/178218971326 \\ 
  7 & Food Science & www.facebook.com/165396023578703 \\ 
  8 & Food Science and Nutrition & www.facebook.com/117931493622 \\ 
  9 & I fucking love science & www.facebook.com/367116489976035 \\ 
  10 & LiveScience & www.facebook.com/30478646760 \\ 
  11 & Medical Laboratory Science & www.facebook.com/122670427760880 \\ 
  12 & National Geographic Magazine & www.facebook.com/72996268335 \\ 
  13 & National Science Foundation (NSF) & www.facebook.com/30037047899 \\ 
  14 & Nature & www.facebook.com/6115848166 \\ 
  15 & Nature Education & www.facebook.com/109424643283 \\ 
  16 & Nature Reviews & www.facebook.com/328116510545096 \\ 
  17 & News from Science & www.facebook.com/100864590107 \\ 
  18 & Popular Science & www.facebook.com/60342206410 \\ 
  19 & RealClearScience & www.facebook.com/122453341144402 \\ 
  20 & Science & www.facebook.com/96191425588 \\ 
  21 & Science and Mathematics & www.facebook.com/149102251852371 \\ 
  22 & Science Channel & www.facebook.com/14391502916 \\ 
  23 & Science Friday & www.facebook.com/10862798402 \\ 
  24 & Science News Magazine & www.facebook.com/35695491869 \\ 
  25 & Science-Based Medicine & www.facebook.com/354768227983392 \\ 
  26 & Science-fact & www.facebook.com/167184886633926 \\ 
  27 & Science, Critical Thinking and Skepticism & www.facebook.com/274760745963769 \\ 
  28 & Science: The Magic of Reality & www.facebook.com/253023781481792 \\ 
  29 & ScienceDaily & www.facebook.com/60510727180 \\ 
  30 & ScienceDump & www.facebook.com/111815475513565 \\ 
  31 & ScienceInsider & www.facebook.com/160971773939586 \\ 
  32 & Scientific American magazine & www.facebook.com/22297920245 \\ 
  33 & Scientific Reports & www.facebook.com/143076299093134 \\ 
  34 & Sense About Science & www.facebook.com/182689751780179 \\ 
  35 & Skeptical Science & www.facebook.com/317015763334 \\ 
  36 & The Beauty of Science \& Reality. & www.facebook.com/215021375271374 \\ 
  37 & The Flame Challenge & www.facebook.com/299969013403575 \\ 
  38 & The New York Times - Science & www.facebook.com/105307012882667 \\ 
  39 & Wired Science & www.facebook.com/6607338526 \\ 
  40 & All Science, All the Time & www.facebook.com/247817072005099 \\ 
  41 & Life's Little Mysteries & www.facebook.com/373856446287 \\ 
  42 & Reason Magazine & www.facebook.com/17548474116 \\ 
  43 & Nature News and Comment & www.facebook.com/139267936143724 \\ 
  44 & Astronomy Magazine & www.facebook.com/108218329601 \\ 
  45 & CERN & www.facebook.com/169005736520113 \\ 
  46 & Citizen Science & www.facebook.com/200725956684695 \\ 
  47 & Cosmos & www.facebook.com/143870639031920 \\ 
  48 & Discover Magazine & www.facebook.com/9045517075 \\ 
  49 & Discovery News & www.facebook.com/107124643386 \\ 
  50 & Genetics and Genomics & www.facebook.com/459858430718215 \\ 
  51 & Genetic Research Group & www.facebook.com/193134710731208 \\ 
  52 & Medical Daily & www.facebook.com/189874081082249 \\ 
  53 & MIT Technology Review & www.facebook.com/17043549797 \\ 
  54 & NASA - National Aeronautics and Space Administration & www.facebook.com/54971236771 \\ 
  55 & New Scientist & www.facebook.com/235877164588 \\ 
  56 & Science Babe & www.facebook.com/492861780850602 \\ 
  57 & ScienceBlogs & www.facebook.com/256321580087 \\ 
  58 & Science, History, Exploration & www.facebook.com/174143646109353 \\ 
  59 & Science News for Students & www.facebook.com/136673493023607 \\ 
  60 & The Skeptics Society \& Skeptic Magazine & www.facebook.com/23479859352 \\ 
  61 & Compound Interest & www.facebook.com/1426695400897512 \\ 
  62 & Kevin M. Folta & www.facebook.com/712124122199236 \\ 
  63 & Southern Fried Science & www.facebook.com/411969035092 \\ 
  64 & ThatsNonsense.com & www.facebook.com/107149055980624 \\ 
  65 & Science \& Reason & www.facebook.com/159797170698491 \\ 
  66 & ScienceAlert & www.facebook.com/7557552517 \\ 
  67 & Discovery & www.facebook.com/6002238585 \\ 
  68 & Critical Thinker Academy & www.facebook.com/175658485789832 \\ 
  69 & Critical Thinking and Logic Courses in US Core Public School Curriculum & www.facebook.com/171842589538247 \\ 
  70 & Cultural Cognition Project & www.facebook.com/287319338042474 \\ 
  71 & Foundation for Critical Thinking & www.facebook.com/56761578230 \\ 
  72 & Immunization Action Coalition & www.facebook.com/456742707709399 \\ 
  73 & James Randi Educational Foundation & www.facebook.com/340406508527 \\ 
  74 & NCSE: The National Center for Science Education & www.facebook.com/185362080579 \\ 
  75 & Neil deGrasse Tyson & www.facebook.com/7720276612 \\ 
  76 & Science, Mother Fucker. Science & www.facebook.com/228620660672248 \\ 
  77 & The Immunization Partnership & www.facebook.com/218891728752 \\ 
  78 & Farm Babe & www.facebook.com/1491945694421203 \\ 
  79 & Phys.org & www.facebook.com/47849178041 \\ 
  80 & Technology Org & www.facebook.com/218038858333420 \\ 
  81 & Biology Fortified, Inc. & www.facebook.com/179017932138240 \\ 
  82 & The Annenberg Public Policy Center of the University of Pennsylvania & www.facebook.com/123413357705549 \\ 
  83 & Best Food Facts & www.facebook.com/200562936624790 \\ 
   \hline
\end{longtable}
\end{footnotesize}

\newpage
\subsection*{Debunking pages}

\begin{footnotesize}
\begin{longtable}{|l|l|l|}
  \hline
  & \textbf{Page name} &  \textbf{Page Link}\\ 
  \hline
1 & Refutations to Anti-Vaccine Memes & www.facebook.com/414643305272351 \\ 
  2 & Boycott Organic & www.facebook.com/1415898565330025 \\ 
  3 & Contrails and Chemtrails:The truth behind the myth & www.facebook.com/391450627601206 \\ 
  4 & Contrail Science & www.facebook.com/339553572770902 \\ 
  5 & Contrail Science and Facts - Stop the Fear Campaign & www.facebook.com/344100572354341 \\ 
  6 & Debunking Denialism & www.facebook.com/321539551292979 \\ 
  7 & The Farmer's Daughter & www.facebook.com/350270581699871 \\ 
  8 & GMO Answers & www.facebook.com/477352609019085 \\ 
  9 & The Hawaii Farmer's Daughter & www.facebook.com/660617173949316 \\ 
  10 & People for factual GMO truths (pro-GMO) & www.facebook.com/255945427857439 \\ 
  11 & The Questionist & www.facebook.com/415335941857289 \\ 
  12 & Scientific skepticism & www.facebook.com/570668942967053 \\ 
  13 & The Skeptic's Dictionary & www.facebook.com/195265446870 \\ 
  14 & Stop the Anti-Science Movement & www.facebook.com/1402181230021857 \\ 
  15 & The Thinking Person's Guide to Autism & www.facebook.com/119870308054305 \\ 
  16 & Antiviral & www.facebook.com/326412844183079 \\ 
  17 & Center for Inquiry & www.facebook.com/5945034772 \\ 
  18 & The Committee for Skeptical Inquiry & www.facebook.com/50659619036 \\ 
  19 & Doubtful News & www.facebook.com/283777734966177 \\ 
  20 & Hoax-Slayer & www.facebook.com/69502133435 \\ 
  21 & I fucking hate pseudoscience & www.facebook.com/163735987107605 \\ 
  22 & The Genetic Literacy Project & www.facebook.com/126936247426054 \\ 
  23 & Making Sense of Fluoride & www.facebook.com/549091551795860 \\ 
  24 & Metabunk & www.facebook.com/178975622126946 \\ 
  25 & Point of Inquiry & www.facebook.com/32152655601 \\ 
  26 & Quackwatch & www.facebook.com/220319368131898 \\ 
  27 & Rationalwiki & www.facebook.com/226614404019306 \\ 
  28 & Science-Based Pharmacy & www.facebook.com/141250142707983 \\ 
  29 & Skeptical Inquirer & www.facebook.com/55675557620 \\ 
  30 & Skeptic North & www.facebook.com/141205274247 \\ 
  31 & The Skeptics' Guide to the Universe & www.facebook.com/16599501604 \\ 
  32 & Society for Science-Based Medicine & www.facebook.com/552269441534959 \\ 
  33 & Things anti-vaxers say & www.facebook.com/656716804343725 \\ 
  34 & This Week in Pseudoscience & www.facebook.com/485501288225656 \\ 
  35 & Violent metaphors & www.facebook.com/537355189645145 \\ 
  36 & wafflesatnoon.com & www.facebook.com/155026824528163 \\ 
  37 & We Love GMOs and Vaccines & www.facebook.com/1380693538867364 \\ 
  38 & California Immunization Coalition & www.facebook.com/273110136291 \\ 
  39 & Exposing PseudoAstronomy & www.facebook.com/218172464933868 \\ 
  40 & CSICOP & www.facebook.com/157877444419 \\ 
  41 & The Panic Virus & www.facebook.com/102263206510736 \\ 
  42 & The Quackometer & www.facebook.com/331993286821644 \\ 
  43 & Phil Plait & www.facebook.com/251070648641 \\ 
  44 & Science For The Open Minded & www.facebook.com/274363899399265 \\ 
  45 & Skeptic's Toolbox & www.facebook.com/142131352492158 \\ 
  46 & Vaccine Nation & www.facebook.com/1453445781556645 \\ 
  47 & Vaximom & www.facebook.com/340286212731675 \\ 
  48 & Voices for Vaccines & www.facebook.com/279714615481820 \\ 
  49 & Big Organic & www.facebook.com/652647568145937 \\ 
  50 & Chemtrails are NOT real, idiots are. & www.facebook.com/235745389878867 \\ 
  51 & Sluts for Monsanto & www.facebook.com/326598190839084 \\ 
  52 & Stop Homeopathy Plus & www.facebook.com/182042075247396 \\ 
  53 & They Blinded Me with Pseudoscience & www.facebook.com/791793554212187 \\ 
  54 & Pro-Vaccine Shills for Big Pharma, the Illumanati, Reptilians, and the NWO & www.facebook.com/709431502441281 \\ 
  55 & Pilots explain Contrails - and the Chemtrail Hoax & www.facebook.com/367930929968504 \\ 
  56 & The Skeptical Beard & www.facebook.com/325381847652490 \\ 
  57 & The Alliance For Food and Farming & www.facebook.com/401665083177817 \\ 
  58 & Skeptical Raptor & www.facebook.com/522616064482036 \\ 
  59 & Anti-Anti-Vaccine Campaign & www.facebook.com/334891353257708 \\ 
  60 & Informed Citizens Against Vaccination Misinformation & www.facebook.com/144023769075631 \\ 
  61 & Museum of Scientifically Proven Supernatural and Paranormal Phenomena & www.facebook.com/221030544679341 \\ 
  62 & Emergent & www.facebook.com/375919272559739 \\ 
  63 & Green State TV & www.facebook.com/128813933807183 \\ 
  64 & Kavin Senapathy & www.facebook.com/1488134174787224 \\ 
  65 & vactruth.com Exposed & www.facebook.com/1526700274269631 \\ 
  66 & snopes.com & www.facebook.com/241061082705085 \\ 
   \hline
\end{longtable}
\end{footnotesize}

\end{singlespacing}

\newpage
\bibliographystyle{unsrt}
\bibliography{biblio_final}

\end{document}